\documentclass{article} 
\usepackage[final]{colm2026_conference}

\usepackage{microtype}
\usepackage{hyperref}
\usepackage{url}
\usepackage{booktabs}

\usepackage{lineno}

\usepackage{times}
\usepackage{latexsym}
\usepackage{booktabs}   
\usepackage{pifont}  
\usepackage{subcaption}
\usepackage{wrapfig}
\usepackage[T1]{fontenc}

\usepackage[utf8]{inputenc}

\usepackage{microtype}
\usepackage{graphicx}
\usepackage{float}

\usepackage[most]{tcolorbox}
\usepackage{amssymb} 

\usepackage{enumitem}
\usepackage{xcolor}

\usepackage{verbatim}

\usepackage{inconsolata}
\usepackage{multirow}

\definecolor{darkblue}{rgb}{0, 0, 0.5}
\hypersetup{colorlinks=true, citecolor=darkblue, linkcolor=darkblue, urlcolor=darkblue}

\title{Document-as-Image Representations Fall Short for\\Scientific Retrieval}

\author{%
  Ghazal Khalighinejad, Raghuveer Thirukovalluru, 
Alexander H. Oh, Bhuwan Dhingra \\
  \\
  Department of Computer Science\\
  Duke University\\ 
}

\begin{document}

\ifcolmsubmission
\linenumbers
\fi

\maketitle

\begin{abstract}
Previous document embedding models are trained on \emph{document-as-image} representations, in which documents are represented as a collection of page images. Document retrieval benchmarks such as ArXivQA and ViDoRe follow the same format, which may favor models trained on document images. However, we argue that this approach is less suitable for text-rich multimodal scientific documents, where relevant information is spread across text, tables, and figures. To study this setting, we introduce a benchmark, arXivDoc, which is constructed from the LaTeX sources of scientific papers from arXiv. These latex sources provide direct access to the structured elements (e.g., sections, tables, figures, equations), which document images represent as pixels. We use these sources to synthetically generate queries about text, tables, and figures, and evaluate text-only, image-only, and multimodal representations using both single-vector and multi-vector retrieval models. Our results show that: (1) document-as-image representations are consistently suboptimal, especially as document length increases; (2) text-based representations are most effective, even for figure-based queries, by using captions and the text around the figure; and (3) interleaved text+image representations outperform document-as-image approaches without training.
\end{abstract}

\begin{figure}[h]
    \centering
    \includegraphics[width=0.76\linewidth]{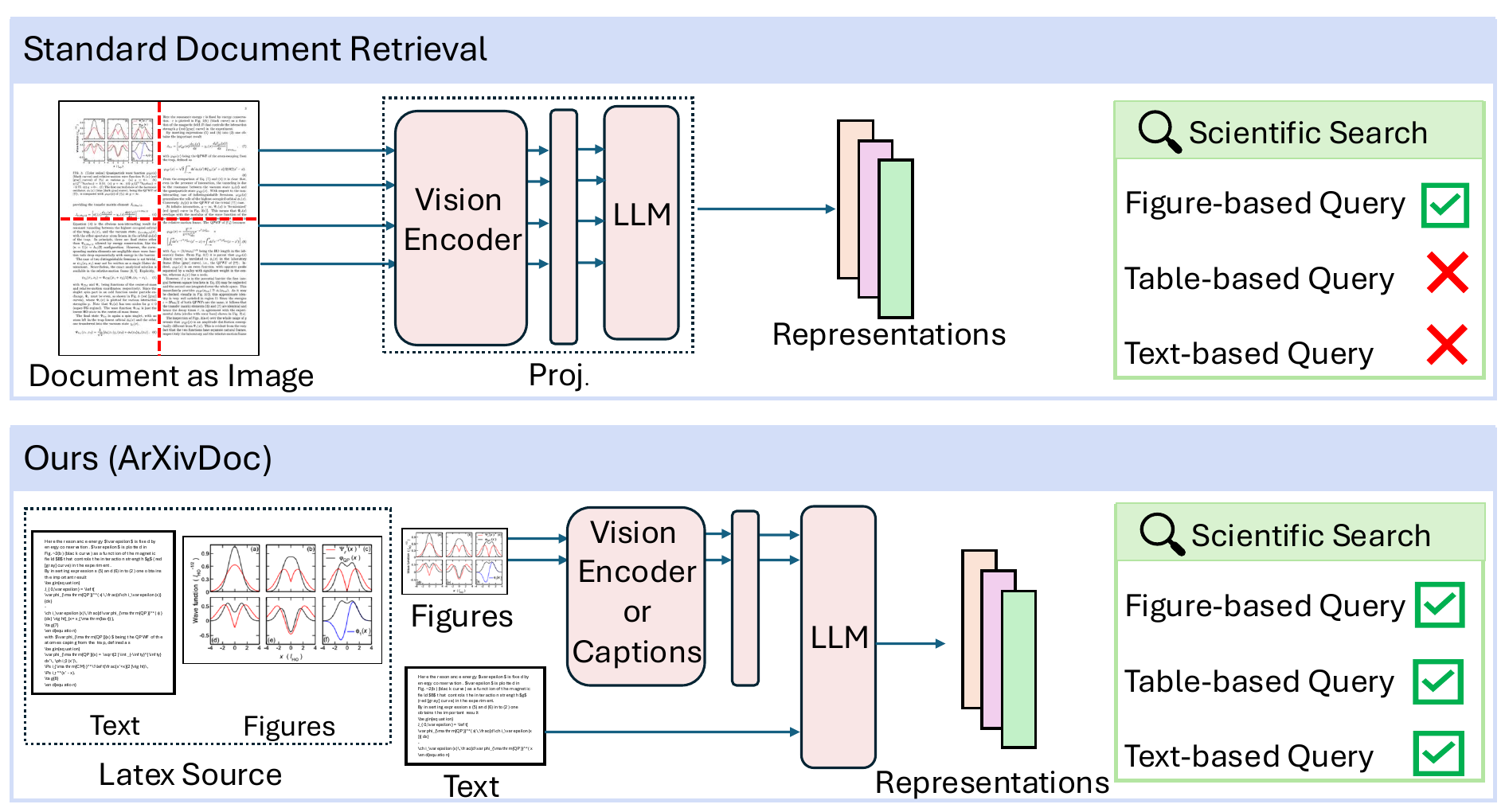}
    \caption{Comparison of document representation paradigms for scientific retrieval. Standard document-as-image approaches process rendered pages through a vision encoder, which handles figure-based queries but struggles to accurately ground text- and table-based evidence. In contrast, arXivDoc, leverages underlying LaTeX source files to process text and figures natively. This preserves fine-grained document structure, enabling robust retrieval across all multimodal query types.
    }
    \label{fig:placeholder}
\end{figure}

\section{Introduction}

Scientific document retrieval requires finding information that may appear in text, equations, tables, or figures. For example, answering a query about a scaling law may require evidence from an equation and its surrounding discussion, while answering a query about model architecture may require information from a figure. Most existing retrieval systems represent documents either as text (e.g., Qwen3-Embedding~\citep{qwen3embedding}, Llama-Embed~\citep{babakhin2025llamaembednemotron8buniversaltextembedding}) or as images (e.g., ColPali~\citep{faysse2024colpali}, Qwen-VL-Embedding~\citep{li2026qwen3vlembeddingqwen3vlrerankerunifiedframework}). However, it remains unclear how representing text and tables as images affects retrieval performance, especially as the text portion gets longer. Therefore, in this paper, we address the following question: \emph{Which document representations are most appropriate for scientific retrieval, and under what conditions?}

Recent work has used document-as-image representations for embedding, following the success of vision–language models (VLMs) that encode rendered pages into dense vectors~\citep{visrag, ma-etal-2024-unifying, faysse2024colpali, vidorev2, gunther-etal-2025-jina, 11093150}. But many scientific papers, particularly in fields that rely on equations, are written in LaTeX, which preserves document structure and content types. Converting these papers into images means that this information is now encoded as pixels and models must infer boundaries between text, equations, and figures from pixels, while cross-references and other structural information are no longer explicitly available. As documents contain more text, it is unclear how well these image-based representations perform. This is particularly relevant for scientific papers, where evidence is often found in text, equations, and tables rather than figures. This concern is also reflected in previous work. PixelWorld shows that pixel-based inputs struggle with text-centric reasoning tasks, while other studies have found that VLM performance on scientific documents depends on the layout and rendering template~\citep{lyu2025pixelworldfarperceivingpixels,cheng2025glyph}.

There is an inherent trade-off between textual and visual document representations. Text-based representations are closer to the data used in pretraining and preserve information such as terminology, equations, and document structure. Image-based representations, on the other hand, preserve layout and figures but require models to infer content boundaries and relationships from images~\citep{faysse2024colpali, wei2025deepseekocrcontextsopticalcompression, lyu2025pixelworldfarperceivingpixels,jiang2024e5}. Visual representations are particularly useful for documents where reliable text extraction or structured sources are unavailable~\citep{vidorev2, mmlongbench-doc, Cho2024M3DocRAG}, such as infographics and slides. However, many scientific papers are available in LaTeX or XML formats that preserve both content and structure. Our text-based representations also work when these sources are unavailable. For papers available only as PDFs, we extract text using MinerU~\citep{wang2026mineru2} or PyMuPDF~\citep{pymupdf}. Replacing LaTeX-derived text with the extracted text does not change our main findings.

In this work, we introduce \textbf{arXivDoc}, a benchmark for analyzing scientific document retrieval across representations. The benchmark consists of a corpus of $8,210$ documents with $144,653$ pages, and $547$ synthetic queries that are manually verified. We consider multiple document representations—text-only, figure-only, text + VLM captions, document-as-image, and interleaved text--image. Each query is explicitly grounded in text, tables, or figures. Our main findings are as follows.
\begin{enumerate}
\item \textbf{Document-as-image is consistently suboptimal}, even for figure-based queries where image representations might be expected to excel. We highlight that this does not depend on access to LaTeX source: replacing LaTeX-derived text with MinerU or PyMuPDF output changes average nDCG@10 by at most $0.02$, with text-based representations remaining superior across all query types.
\item \textbf{Text + VLM captions achieves the strongest overall performance}, demonstrating that augmenting text with vision--language model descriptions of figures and tables is more effective than replacing text with images.
\item \textbf{Interleaved text--image representations outperform document-as-image.} We reuse the same embedding models that were originally trained on page-level document images, and apply them to interleaved text--image inputs without any additional training. Despite this mismatch, they still outperform document-as-image representations.
\item \textbf{As more text is added to a page, document-as-image representations degrade more sharply than text-based representations.} This makes text-based representations more suitable for long scientific documents. 
\end{enumerate}

\begin{table}[h]
\centering
\small
\resizebox{\textwidth}{!}{%
\begin{tabular}{lrrrrr|cc|ccc}
\toprule
\textbf{Benchmark} & \textbf{\# Docs} & \textbf{\# Pages} & \textbf{\# Queries} & \textbf{\# Tokens/Page} & \textbf{Target Unit} & \textbf{Modality} & \textbf{Source Availability} & \textbf{Open-Domain} & \textbf{Scientific} \\
\midrule
QASper & 1{,}585 & -- & 5{,}049 &  -- & Document & Text & \ding{51} & \ding{55} & \ding{51} \\
ArXivQA & 100k & -- & 100k & -- & Figure & Image & \ding{55} & \ding{55} & \ding{51} \\
MRMR & -- & 26,223 & 1{,}435  & 421 & Webpage & Text+Image & \ding{51} & \ding{51} & \ding{51} \\
ViDoRe V1 & -- & -- & 5{,}000  & -- & Page/Figure & Image & \ding{55} & \ding{55} & \ding{55} \\
ViDoRe V2 & 66 & 3{,}266 & 3{,}000  & 488 & Page & Image & \ding{55} & \ding{51} & \ding{51} \\
NL-DIR & -- & 41{,}000 & 205k  & -- & Page & Image & \ding{55} & \ding{55} & \ding{55} \\
MMDocIR & 313 & -- & 1{,}658  & 700 & Page & Text+Image & \ding{55} & \ding{55} & \ding{51} \\
\midrule
\textbf{arXivDoc} (\textbf{Ours}) & \textbf{8{,}210} & 144,653 & \textbf{547}  & 947 & \textbf{Document} & \textbf{Text+Image} & \ding{51} & \ding{51} & \ding{51} \\
\bottomrule
\end{tabular}%
}
\caption{Comparison of document-level multimodal and scientific retrieval benchmarks. \textbf{Target Unit} indicates the expected retrieval granularity (e.g., retrieving a specific page vs. the entire document). \textbf{Modality} indicates the default representation of the corpus. ArXivDoc is the only large-scale, open-domain scientific retrieval benchmark providing raw LaTeX source access alongside rendered images.}
\label{tab:dataset-comparison-with-queries}
\end{table}
\section{Related Work}
In this section, we position arXivDoc within existing document retrieval and understanding benchmarks, highlighting the key gaps that remain. Table \ref{tab:dataset-comparison-with-queries} summarizes these differences.

ArXivDoc is an \emph{open-domain scientific document retrieval} task: given a query, the model must retrieve the single relevant paper from a corpus of more than $8{,}000$ arXiv documents. Queries are context-independent and grounded in specific evidence (text, tables, or figures). This differs from benchmarks such as ArXivQA~\citep{li-etal-2024-multimodal-arxiv}, which focus on retrieving from a pool of figures rather than complete multimodal documents.

Several document benchmarks do not operate in a true retrieval setting. For instance, QASPER~\citep{dasigi2021qasper} assumes the document is given and evaluates question answering for queries within a paper, while MMLongBench-Doc~\citep{mmlongbench-doc}, MMDocIR~\citep{dong-etal-2025-mmdocir}, and UniDoc-Bench~\citep{peng2026unidocbenchunifiedbenchmarkdocumentcentric} focus on restricted or non-open-domain queries where the relevant context is already provided. As a result, queries do not require identifying the correct document from a large corpus, and models are not challenged to distinguish among many candidates—sidestepping the central difficulty of scientific retrieval.

Other benchmarks evaluate retrieval at the \emph{page level} using document images. ViDoRe V1 and V2~\citep{faysse2024colpali,vidorev2} retrieve relevant pages from visually rich documents, and NL-DIR~\citep{guo2025towards} extends this paradigm to larger-scale image-based retrieval. In these settings, documents are decomposed into pages rather than treated as unified documents. While effective for visually salient content, this formulation is less suitable for scientific papers, where evidence is often distributed across text, equations, tables, and figures spanning multiple pages. Additionally, while OCR can be applied to extract text or isolate elements such as figures and tables, it requires additional preprocessing~\citep{choi-etal-2025-zero,han2025mdocagent} that is often error-prone and computationally expensive, introducing noise and variability in the extracted content. As a result, it becomes difficult to systematically evaluate how different document representations (e.g., text vs.\ image) impact retrieval performance.

MRMR~\citep{mrmr} is closer in spirit to our setting in that it studies open-domain retrieval over multimodal documents with access to structured sources. However, the underlying document distribution differs substantially: MRMR operates over webpages, which are typically less information-dense and less structurally explicit than scientific papers, where critical evidence is often embedded in tightly coupled text, equations, tables, and figures. More importantly, the task objectives diverge. ArXivDoc is designed to reflect a realistic scientific search scenario, where a natural, context-independent query must be used to identify the single correct paper from a large corpus of over $8{,}000$ documents. In contrast, MRMR emphasizes reasoning-intensive retrieval, where queries are often highly specific and tied to localized webpage content, frequently requiring deep multimodal interpretation (e.g., understanding an image) to resolve.

ArXivDoc addresses these gaps. It combines: (1) \emph{open-domain retrieval}, (2) \emph{scientific documents}, and (3) \emph{document-level retrieval}. In addition, it provides access to the underlying document sources (e.g., LaTeX when available) alongside rendered pages, which allows us to compare text-based and page-based representations directly. Finally, queries are grounded in specific evidence types (text, tables, figures), enabling controlled evaluation across modalities.

\begin{figure*}[t] 
    \centering
    \includegraphics[width=\textwidth]{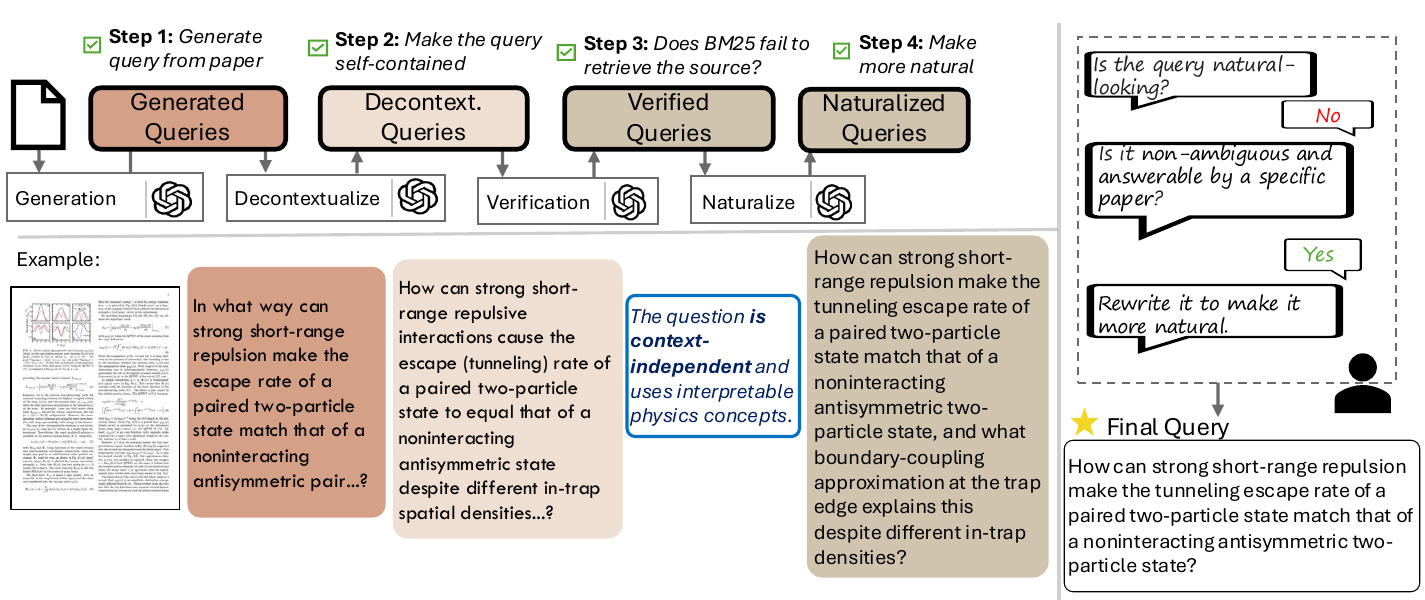}
    \caption{Dataset construction pipeline. Query counts across stages: LLM verification reduces queries from 761→525 (text), 650→159 (figure), and 2648→280 (table); manual verification further reduces them to 229 (text), 100 (figure), and 218 (table).
    }
    \label{fig:full_width_image}
\end{figure*}

\section{Dataset}

\begin{wraptable}{r}{0.32\linewidth}
\centering
\scriptsize

\textbf{Page distribution}
\vspace{1pt}

\begin{tabular}{lrr}
\toprule
\textbf{Range} & \textbf{Count} & \textbf{\%} \\
\midrule
1--5 pages   & 730   & 8.9\%  \\
6--10 pages  & 2,170 & 26.4\% \\
11--20 pages & 3,152 & 38.4\% \\
21--40 pages & 1,701 & 20.7\% \\
41+ pages    & 455   & 5.5\%  \\
\bottomrule
\end{tabular}

\vspace{3pt}

\textbf{Token distribution}
\vspace{1pt}

\begin{tabular}{lrr}
\toprule
\textbf{Range} & \textbf{Count} & \textbf{\%} \\
\midrule
$<$1k    & 20    & 0.2\% \\
1k--5k   & 300   & 3.7\% \\
5k--10k  & 2,429 & 29.6\% \\
10k--20k & 3,317 & 40.4\% \\
20k+     & 2,144 & 26.1\% \\
\bottomrule
\end{tabular}

\caption{Page and token distributions}
\end{wraptable}

We introduce \textit{arXivDoc}, an open-domain scientific document retrieval benchmark constructed from raw LaTeX sources. We use the LaTeX sources to generate queries and construct document representations, rather than extracting text from PDFs using OCR. This lets us preserve the document content and structure while comparing different representations of the same papers.

The benchmark includes 8{,}210 full-length scientific papers, with substantially more pages than previous multimodal retrieval benchmarks (Table~\ref{tab:dataset-comparison-with-queries}). Many of these papers cover similar topics and share notation and experimental setups, making retrieval more challenging. Examples of documents and queries are provided in Appendix~\ref{app:query_examples}.

Queries are generated, filtered, and verified through a multi-stage pipeline using \texttt{gpt-5.2}, and are then manually rewritten, edited, and validated by human annotators. In total, the dataset contains $547$ queries: $100$ figure-based, $218$ table-based, and $229$ text-based queries.

\paragraph{Design Trade-offs in Query Construction.}
Retrieval benchmark design involves a fundamental trade-off between three competing objectives: 
(i) naturalness of queries, 
(ii) decontextualization (self-contained queries that do not rely on implicit document references), and 
(iii) unambiguous ground truth. 

In practice, these objectives are often in tension. Natural queries tend to be short and underspecified, relying on implicit context and permitting multiple valid answers, which increases ambiguity and leads to false negatives. This issue is evident in existing datasets such as ArXivQA~\citep{li-etal-2024-multimodal-arxiv}, MMDocIR~\citep{dong-etal-2025-mmdocir}, and ViDoRe V1~\citep{faysse2024colpali}, where queries are frequently ambiguous and admit multiple valid matches, making them less suitable for document retrieval (e.g., \textit{“What is plotted along the x axis?”} from ViDoRe can correspond to many documents). We also identified multiple false negatives in ViDoRe V2: despite its framing as blind contextual querying, some queries remain close-domain or admit multiple plausible matches. 

Other benchmarks avoid this problem by enforcing decontextualization during dataset construction; for example, MultimodalQA~\citep{talmormultimodalqa, Cho2024M3DocRAG} uses formal languages to generate cross-modal queries. However, such design choices can come at the expense of naturalness. ArXivDoc instead prioritizes decontextualization and unambiguous ground truth while keeping queries as natural as possible, enabling controlled evaluation of open-domain scientific document retrieval over a large corpus, where the goal is to identify the correct document among many topically similar papers. In this setting, minimizing ambiguity is essential for reliable evaluation. Table~\ref{tab:query_examples} shows examples of queries from different datasets.

\begin{table}[t]
\centering
\small
\setlength{\tabcolsep}{1pt}
\begin{tabular}{p{3cm}p{10cm}}
\toprule
\textbf{Dataset} & \textbf{Example query}  \\
\midrule
ArXivQA
& \textit{What does the black sphere with an arrow represent in these diagrams?} \\

ViDoRe V1
& \textit{What process is being depicted in the figure?} \\

MMDocIR
& \textit{What variable is being altered to create the different patterns in each subplot?}\\

\textbf{arXivDoc}
& \textit{Why do UV-based measurements of the low-redshift star formation rate density often come out higher than other estimates?} \\
\bottomrule
\end{tabular}
\caption{Example queries from existing benchmarks and arXivDoc. Existing benchmarks often contain short queries that depend on the surrounding context. Unlike these, arXivDoc uses context-independent queries.}
\label{tab:query_examples}
\end{table}
\subsection{Query Generation and Filtering}
\label{sec:query_generation}

We generate open-domain retrieval queries from three types of evidence in scientific papers: \emph{text}, \emph{tables}, and \emph{figures}. We create a pool of candidate queries for each type and apply the same generation and filtering steps to all three.

For \textbf{text-based} queries, we use content from LaTeX (\texttt{.tex}) files. \textbf{Figure-based} queries are generated from figures in the LaTeX sources, and \textbf{table-based} queries from \texttt{\textbackslash begin{table}} environments. Each query goes through several steps: synthetic generation, decontextualization, difficulty-based filtering, verification, and naturalization (Figure~\ref{fig:full_width_image}). The resulting queries are then reviewed by human annotators, who edit or remove queries that do not meet our criteria (See Appendix~\ref{app:query_generation} for the prompts used at each stage and the annotation instructions).

We highlight that queries in arXivDoc are grounded in localized evidence (a specific text span, table, or figure) rather than requiring multi-hop reasoning across different sections. This is intentional because we intend to focus on document retrieval rather than document-level reasoning: the primary challenge is identifying the correct document from a large corpus, not aggregating evidence within a document.

\paragraph{Generation}
For each type of source (text, table, or figure), we prompt \texttt{gpt-5.2} to generate a query that targets the underlying scientific content. We instruct the model to generate queries that require expert-level reasoning about implications, trends, limitations, or constraints. The queries should avoid directly repeating the source text and instead use paraphrasing to reduce lexical overlap. They should also be answerable from the document without relying on keyword matching or references to specific sections, figures, or experiment names. Each query must be a single, concise, and realistic sentence. Full prompt templates are provided in Appendix~\ref{app:query_generation}.

\paragraph{Decontextualization.}
We find that many generated queries depend on the context. For example, the query
\emph{Can the top polyhedron be obtained from the cube by a shear?''}
refers to a specific figure and is not self-contained. We use \texttt{gpt-5.2} to rewrite these queries so they can be understood without the original figure or table. In this example, the query is rewritten as
\emph{Can an oblique parallelepiped be obtained from a cube by an affine shear?''},
which replaces the figure-specific reference with geometric terminology.

However, we find that the model often makes only minor changes to the original query without adding enough context. For example:

\vspace{0.5em}
\noindent\textbf{Original:} \emph{If the variable on the x-axis represents time, what can be inferred about the rate of change of the parameter over time?}

\noindent\textbf{Rewritten:} \emph{If the independent variable represents time, what can be inferred about the rate of change of the parameter over time?}

\vspace{0.5em}

The rewritten query still does not provide enough information to be understood without the original figure. Although we instruct \texttt{gpt-5.2} to return \texttt{null} when it cannot produce a valid rewrite, it frequently returns small paraphrases like this instead. We therefore add a final verification step to identify queries that remain context-dependent.

\paragraph{Difficulty-Based Filtering.}
We use BM25~\citep{bm25s} to filter out queries that are too easy to retrieve through lexical matching. For each query, we run BM25 over a chunked document corpus. If the gold document appears among the top five results, we remove the query, since this suggests that simple lexical matching may be sufficient to retrieve it. This removes roughly $40\%$ of the queries.

\paragraph{Final Verification.}
We use \texttt{gpt-5.2} to check whether the remaining queries can be understood without the original document. Queries that still depend on document-specific context are removed. After this step, the number of queries decreases from $761$ to $525$ for text, $650$ to $159$ for figures, and $2648$ to $280$ for tables.

\paragraph{Human Annotation}
We manually review the queries for \textit{naturalness}, \textit{ambiguity}, and \textit{document answerability}. Queries that are unclear, lack sufficient context, or cannot be answered using evidence from the gold document are edited or removed. Three annotators participate in this process: two PhD students and one Master's student. The Master's student reviews all queries to check whether they are clear and plausible (naturalness), identify a unique target document (ambiguity), and are supported by evidence in the gold document (document answerability). See annotation details in Appendix~\ref{fig:human_annotation}.
\section{Experiments}

\subsection{Problem Formulation}
\label{sec:problem_formulation}

We study open-domain scientific document retrieval. Let \(D = \{d_1, \dots, d_N\}\) be a corpus of scientific documents. Each document \(d_i\) is represented by a set of embedding units,
\[
\mathbf{e}_i = \{e_{i1}, \dots, e_{iM_i}\},
\]
where \(e_{ij}\) is the \(j\)-th unit of document \(d_i\). Depending on the representation, a unit may be a text chunk, figure, or page. Given a context-independent query \(q\) about a specific piece of scientific evidence (text, table, or figure), we score each unit with \(s(q, e_{ij})\). The document score is the maximum score over its units, \(S(q, d_i) = \max_{e_{ij} \in \mathbf{e}_i} s(q, e_{ij})\). We rank documents using this score.

\subsection{Experimental Setup}
We use \textbf{arXivDoc} to compare document representations for open-domain retrieval from scientific papers. Starting from the LaTeX sources, we construct different representations of the same documents and measure retrieval performance and index size. We report normalized discounted cumulative gain at rank 10 (nDCG@10), which measures how highly the relevant document is ranked.

\paragraph{Representations and Models.}
We consider three classes of document representations: \textbf{text-only}, \textbf{image-only}, and \textbf{text + image}. Within each class, we evaluate multiple representations.

\textbf{Text-only.}
(i) \textbf{Text (\LaTeX)}, which indexes raw \LaTeX{} source text. We first \emph{flatten} the source to a single file to account for projects split across multiple \texttt{.tex} inputs (e.g., via \texttt{\textbackslash input} or \texttt{\textbackslash include}). We then apply lightweight normalization to remove non-semantic markup: comments and common formatting commands (e.g., \texttt{\textbackslash cite}, \texttt{\textbackslash ref}, \texttt{\textbackslash label}, \texttt{\textbackslash footnote}, and styling macros such as \texttt{\textbackslash emph}, \texttt{\textbackslash textbf}), while preserving scientific content such as plain text, math, and structure. (ii) \textbf{Text + VLM Captions}, which augments document text with figure descriptions generated by a vision--language model, appended to the end of the document. Caption-generation details, along with a sensitivity analysis showing that results are robust to the choice of captioning model and prompt, are provided in Appendix~\ref{app:caption_sensitivity}.

\textbf{Image-only.}
(iii) \textbf{\LaTeX{} Figures}, which indexes rendered figures extracted from the \LaTeX{} sources while ignoring document text. We collect all figure assets (e.g., \texttt{.png}, \texttt{.jpg}, \texttt{.pdf}, \texttt{.eps}) and convert them into a unified format prior to embedding. This representation isolates visual content, but may miss critical information when the evidence required to answer a query resides in the text.

\textbf{Text + Image.}
(iv) \textbf{Document-as-Image}, which indexes full document pages rendered as images. (v) \textbf{Interleaved Text + Images}, which uses both text and figures in their original order. We extract text and figure references from the \LaTeX{} source, resolve the references to rendered images, and arrange the content as it appears in the paper. We then split the text into chunks and combine each chunk with nearby figures to create multimodal retrieval units containing one or two images.

Across these representations, we evaluate several embedding models, depending on
modality compatibility: \textbf{Qwen3-Embedding-8B}~\citep{qwen3embedding} as a
text-only embedder, and \textbf{Qwen3-VL-Embedding-8B}~\citep{li2026qwen3vlembeddingqwen3vlrerankerunifiedframework},
\textbf{ColQwen2 v1} and \textbf{ColQwen2.5-v0.2}~\citep{faysse2024colpali},
\textbf{OpenCLIP ViT-G/14}~\citep{cherti2023reproducible,radford2021learning}, and
\textbf{Ops-MM-Embedding v1}~\citep{linmm} as image embedders. For each
representation--model pair, we report results using the best-performing configuration;
full hyperparameter sweeps are reported separately.

ColQwen is a late-interaction model based on the ColBERT framework~\citep{khattab2020colbertefficienteffectivepassage, santhanam-etal-2022-colbertv2} and built on Qwen2-VL~\citep{wang2024qwen2vl}. It encodes queries and documents into sets of token-level embeddings and computes similarity using max-similarity aggregation. Since it stores multiple embeddings per input, its index is significantly larger than those of single-vector models. In comparison, Qwen, OpenCLIP, and Ops-MM-Embedding produce one embedding per input unit and use standard vector similarity for retrieval. ColQwen was originally developed for page-level visual document retrieval and was not trained on text documents. We still test it on text chunks, where each chunk is an input unit. This lets us compare ColQwen with single-vector text embedding models on the same document inputs. We find that ColQwen performs well on text, even without being trained for this setting. Based on this observation, we also test ColQwen on interleaved text--image representations and find that they outperform document-as-image inputs using the same model. We observe a similar result with Qwen-VL-Embedding, which was also not specifically trained for text embedding. It achieves higher retrieval performance with interleaved text + image inputs than with document-as-image representations.

\paragraph{Index Size.}
Index size is defined as the total storage used for all embedding vectors, except for model parameters. For \textbf{text-based representations}, index size is controlled by varying the \emph{chunk size} used to segment documents. Smaller chunks increase the number of units and the total index size, while larger chunks reduce storage at the cost of coarser representations. This mechanism is used for all text indexings. For ColQwen, varying chunk size does not substantially change the index size, since text is encoded at the token level and stored as a set of embeddings regardless of chunk boundaries.

For vision-based representations, index size is controlled via the \texttt{max\_pixels} parameter, which caps the total number of input pixels processed per image. Images are resized to approximately preserve aspect ratio while satisfying this budget. The number of visual tokens scales with the effective image resolution and can be approximated as \(T_{\text{vis}} \propto \frac{H' W'}{P^2}\), where \(H' W' \le \texttt{max\_pixels}\) is the resized image resolution and \(P\) is the vision encoder’s patch size. Reducing \texttt{max\_pixels} therefore decreases the number of visual tokens and the resulting index size, trading visual detail for storage efficiency. We tune the \texttt{max\_pixels} parameter for each model–representation pair and report the best-performing configuration.

\begin{table}[t]
\centering
\small
\begin{tabular}{l l c r r r r}
\toprule
\textbf{Input} & \textbf{Model} & Index (GB) & Text & Table & Figure & Avg. \\
\midrule

\multirow{4}{*}{Text Only}
 & OpenCLIP & 1.28 & 0.18 & 0.12 & 0.48 & 0.21 \\
 & Ops-MM-Embedding & 2.24 & 0.75 & 0.52 & 0.75 & 0.66 \\
 & ColQwen2 v1 & 40.5 & 0.77 & 0.56 & 0.80 & 0.69 \\
 & ColQwen2.5-v0.2 &  57.1 & 0.77 & 0.59 & 0.80 & 0.70 \\
 & Qwen3-Embedding & 2.56 & 0.87 & 0.60 & 0.76 & 0.74 \\

\midrule

\multirow{4}{*}{Text + VLM Captions}
 & OpenCLIP & 1.6 & 0.19 & 0.11 & 0.50 & 0.21 \\
 & Ops-MM-Embedding & 2.8 & 0.74 & 0.50 & 0.75 & 0.65 \\
 & ColQwen2 v1 & 52.1 & 0.74 & 0.54 & 0.90 & 0.69 \\
  & ColQwen2.5-v0.2 &  67.4 & 0.77 & 0.59 & 0.85 & 0.70 \\
 & Qwen3-Embedding & 3.1 & 0.87 & 0.60 & 0.80 & 0.75 \\
\midrule

\multirow{4}{*}{Figures Only}
 & OpenCLIP & 0.5 & 0.03 & 0.07 & 0.60 & 0.15 \\
 & Ops-MM-Embedding & 0.8 & 0.26 & 0.25 & 0.80 & 0.36 \\
 & ColQwen2 v1 & 35.4 & 0.18 & 0.18 & 0.87 & 0.30 \\
  & ColQwen2.5-v0.2 & 34.6 & 0.19 & 0.19 & 0.87 & 0.31 \\
 & Qwen3-VL-Embedding & 1.0 & 0.22 & 0.25 & 0.83 & 0.35 \\

\midrule

\multirow{4}{*}{Doc-as-Image}
 & OpenCLIP & 0.6 & 0.02 & 0.05 & 0.41 & 0.10 \\
 & Ops-MM-Embedding & 1.0 & 0.70 & 0.51 & 0.75 & 0.63 \\
 & ColQwen2 v1 & 70.5 & 0.73 & 0.52 & 0.84 & 0.67 \\
& ColQwen2.5-v0.2 &  47.5 & 0.77 & 0.57 & 0.86& 0.71 \\
 & Qwen3-VL-Embedding & 1.3 & 0.78 & 0.55 & 0.78 & 0.69 \\

\midrule

\multirow{2}{*}{Interleaved (Text + Image)}
 & ColQwen & 49.5 & 0.78 & 0.56 & 0.82 & 0.70 \\
 & Qwen3-VL-Embedding & 2.9 & 0.85 & 0.57 & 0.75 & 0.72 \\

\bottomrule
\end{tabular}
\caption{Retrieval performance (NDCG@10) across different document representations and models. Index (GB) reflects storage cost.}
\label{tab:main_results}
\end{table}

\subsection{Main Results}

Table~\ref{tab:main_results} shows nDCG@10 for the best configuration of each representation--model pair. Results are separated by text, table, and figure queries. See reported confidence intervals in Appendix~\ref{app:confidence_intervals}.

\paragraph{Main Results.}

\textbf{(1) Document-as-image representations are consistently suboptimal.}
Document-as-image performs worse than alternative representations across text, table, and figure queries. Even on figure-based queries, \textbf{ColQwen2 v1 with text + VLM captions} achieves the highest score using only textual inputs. We also find that \textbf{interleaved text + image} performs better than document-as-image, although both representations contain visual information.

\textbf{(2) Text-only representations are competitive for figure-based queries.}
Text-only retrieval comes close to document-as-image retrieval on figure queries, even without access to the figures or generated captions. Adding VLM captions improves performance further: \textbf{ColQwen2 v1 with text + VLM captions} reaches $0.90$, the highest figure-query score. Scientific papers often describe their figures in captions and nearby text, which may explain why the images themselves are not always needed for retrieval.

\textbf{(3) Interleaved representations outperform document-as-image without dedicated training.}
The embedding models in our experiments were not explicitly trained for interleaved text + image inputs. Even so, interleaving text and figures works better than representing whole pages as images. Keeping related text and figures together appears useful even without additional training for this input format.

\textbf{(4) Chunk-level representations have a better accuracy--storage trade-off in our experiments.}
All methods aggregate scores over multiple embedding units per document, but the units differ in \emph{granularity}. Chunk-level models (Qwen3-Embedding and Qwen3-VL-Embedding) store one vector per chunk or page, while late-interaction models (ColQwen) store one per token or patch. On the same text inputs, Qwen3-Embedding outperforms ColQwen2.5-v0.2 with a ${\sim}22\times$ smaller index. On document-as-image inputs, ColQwen2.5-v0.2 performs slightly better than Qwen3-VL-Embedding but uses ${\sim}37\times$ more storage. This is an accuracy--storage trade-off rather than evidence that one architecture is inherently better. We did not exhaustively test choices such as \texttt{max\_pixels}, token pooling, or dimensionality projection.

\subsection{Do the Conclusions Depend on LaTeX Source?}
\label{sec:parsing}

\begin{table}[t]
\centering
\small
\begin{tabular}{llccccc}
\toprule
Input & Text Source & Model & Text & Table & Figure & Avg. \\
\midrule
\multirow{3}{*}{Text Only}
 & LaTeX   & Qwen3-Embedding & 0.87 & 0.60 & 0.76 & 0.74 \\
 & MinerU  & Qwen3-Embedding & 0.84 & 0.59 & 0.76 & 0.72 \\
 & PyMuPDF & Qwen3-Embedding & 0.85 & 0.59 & 0.77 & 0.73 \\
\midrule
\multirow{3}{*}{Text + VLM Captions}
 & LaTeX   & Qwen3-Embedding & 0.87 & 0.60 & 0.80 & \textbf{0.75} \\
 & MinerU  & Qwen3-Embedding & 0.86 & 0.59 & 0.80 & 0.74 \\
 & PyMuPDF & Qwen3-Embedding & 0.86 & 0.59 & 0.80 & 0.74 \\
\bottomrule
\end{tabular}
\caption{Effect of text-extraction source on retrieval performance (nDCG@10).
Substituting PDF-parsed text for LaTeX source costs at most $0.02$ average nDCG@10,
and both parsed variants remain ahead of document-as-image representations on text and
table queries.}
\label{tab:parsing}
\end{table}

The text-based results in Table~\ref{tab:main_results} use LaTeX sources, but scientific papers are often available only as PDFs. We therefore use text extracted with MinerU~\citep{wang2026mineru2} and PyMuPDF~\citep{pymupdf} for text-based representations. 

Table~\ref{tab:parsing} shows that the results change little. Average nDCG@10 drops by at most $0.02$ relative to LaTeX, with the largest change on text queries and almost no change on table or figure queries. Parsed text remains better than both document-as-image models on text and table queries, and adding captions still gives the strongest results overall.

\subsection{Analysis}

\begin{wrapfigure}{r}{0.5\linewidth}
\includegraphics[width=1\linewidth]{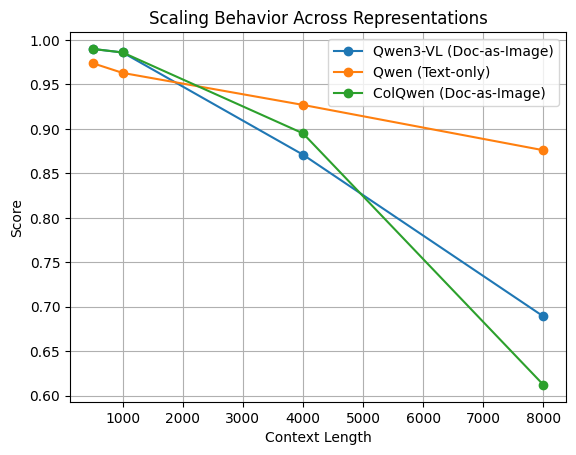}
\caption{
Retrieval performance (NDCG@10) as a function of context length (token length) across document representations. The max-pixels parameters are tuned (see Appendix~\ref{app:hyperparameter}).}
\label{fig:scaling}
\end{wrapfigure}

\subsubsection{Scaling Behavior}
We study how retrieval performance changes as the amount of context increases. Starting from the flattened LaTeX source, we sample a base window of $500$ tokens and progressively expand it to larger contexts ($1000$, $4000$, and $8000$ tokens), while keeping the query fixed across all variants. For each window, we convert the text into a minimally executable LaTeX snippet and render it to a PDF. Then, we extract a single page image. This enables a controlled comparison between text-only and document-as-image representations under matched content. For retrieval, we use a fixed 100-token query sampled from the original 500-token window, from $130$ documents. Figure~\ref{fig:scaling}, shows that document-as-image representations degrade more sharply as context grows, in contrast to text-based representations.

\subsubsection{Why Do Text Representations Work for Figure-Based Queries?}
\label{sec:figure_analysis}

To understand why text representations perform well on figure-based queries, we analyze the highest-scoring retrieved text chunk from the correct document for each query. For each case, we check whether the top chunk (i) lies near the figure, (ii) explicitly references it, or (iii) contains caption-level information. A chunk is considered \emph{near} if it lies within a small window (±2 chunks). To detect figure references, we use regex patterns such as \texttt{Fig.}, \texttt{\textbackslash ref\{fig\}}, and variants of \texttt{Figure X}. To assess caption-level information, we extract informative words from the caption (after removing stopwords) and measure their overlap with the chunk text.

\begin{wraptable}{r}{0.5\linewidth}
\centering
\scriptsize
\setlength{\tabcolsep}{4pt} 
\begin{tabular}{lcc}
\toprule
\textbf{Metric} & \textbf{Correct Doc} & \textbf{Incorrect Doc} \\
\midrule
Count & 63 & 61 \\
\midrule
Near Figure (\%) & 41.3 & 23.0 \\
References Figure (\%) & 60.3 & 42.6 \\
Contains Caption Info (\%) & 68.3 & 24.6 \\
\bottomrule
\end{tabular}
\caption{
Analysis of the highest-scoring retrieved text chunk for figure-based queries.
}
\label{tab:figure_text_analysis}
\end{wraptable} 
As a baseline, we perform the same analysis on top chunks retrieved from incorrect documents. Results are summarized in Table~\ref{tab:figure_text_analysis}. Correct-document retrievals frequently exhibit all three signals, while baseline chunks show lower rates across these indicators. These results suggest that text retrieval does not require access to the figure itself. Instead, it relies on surrounding textual descriptions—references, explanations, and caption content—that encode the figure’s information in text form.

\subsubsection{Comparison with ViDoRe Benchmark}
\label{sec:vidore}

To assess the generalizability of our findings, we evaluate text-only and document-as-image retrieval representations on the ViDoRe benchmark. Since ViDoRe provides only document images, we extract text using PaddleOCR~\citep{cui2025paddleocr} to obtain text-only representations. We then compare text-only representation, against document-as-image, across the subsets.

\begin{wraptable}{r}{0.5\linewidth}
\centering
\scriptsize
\begin{tabular}{lc|c|cc}
\toprule
\multirow{2}{*}{\textbf{Subset}} & \multirow{2}{*}{\textbf{\#Tokens}} & \textbf{Text} & \multicolumn{2}{c}{\textbf{Doc as Image}} \\
\cmidrule(lr){3-3} \cmidrule(lr){4-5} &
 & Qwen & ColQwen & Qwen VL \\
\midrule
Biomedical   &   108  & 0.64 & 0.60 & 0.68 \\
Economics      &  704 & 0.50 & 0.53 & 0.51 \\
ESG     &  582 & 0.50 & 0.56 & 0.65 \\
ESG HL   &   583 & 0.61 & 0.60 & 0.68 \\
\bottomrule
\end{tabular}
\caption{nDCG@10 on ViDoRe dataset.}
\label{tab:ViDoRe}
\end{wraptable}

As shown in Table~\ref{tab:ViDoRe}, the results on ViDoRe differ from those on arxivDoc, with document-as-image representations generally performing better than text-only representations. WWe attribute this discrepancy to two main differences between the benchmarks. First, ArXivDoc contains scientific papers with an average of 947 tokens per page, technical vocabulary, and dense text. ViDoRe, on the other hand, includes lecture slides, corporate ESG reports, and economic reports, which have fewer tokens per page ($108$--$704$) and rely more on layout, charts, and infographics. These differences may explain why visual representations work better on ViDoRe, while text-based representations perform better on scientific papers, where much of the relevant information is found in the text.

Second, ViDoRe's page-level evaluation introduces noise due to the combination of shorter, less specific queries (21 tokens on average, compared to 34 in arxivDoc) and page-level granularity. Because these queries contain less domain-specific terminology, multiple pages from the same document often include relevant information. We sampled $100$ queries where the Qwen text embedder did not retrieve the ground truth in the top $5$ and used an LLM judge (GPT-5.2) to assess the retrieved documents by asking whether the query can be answered given each document. Across the retrieved documents, the judge identified $22\%$ of queries containing at least one missed but answerable result, indicating false negatives, an issue also highlighted by recent works \citep{wang2024mitigating,wasserman2025real}. In contrast, arXivDoc operates at the document level with more specific queries that map cleanly to a single ground truth document, and in our evaluation of $100$ queries we observe no false negatives. These findings indicate that the ViDoRe results should be interpreted with caution.

These findings highlight a key difference between our setting and ViDoRe: document-as-image representations may be advantageous when queries are broad or when multiple pages contain overlapping information, whereas text-based representations benefit from more specific queries and clear document-level grounding. Notably, this distinction is also consistent with findings from the MRMR~\citep{mrmr} study, which similarly reports stronger performance from Qwen3-based text embeddings relative to document-as-image approaches on their document-level dataset.

\section{Conclusion and Future Work}
We introduce \textbf{arXivDoc} benchmark to study how different document representations perform on scientific retrieval task. By working directly with LaTeX sources, we can preserve the text, tables, figures, and equations rather than representing each page as an image. Our experiments show that document-as-image representations generally perform worse than text-based representations, especially on text-heavy scientific papers.

One interesting result is that interleaving text and images works better than document-as-image, even when the embedding models have not been trained on this type of input. It would be interesting to see whether training these models on interleaved scientific documents improves retrieval further. arXivDoc currently focuses on queries grounded in a single text span, table, or figure. We leave queries that require combining information from different parts of a paper, or across several papers, for future work.
\section{Limitations}
\label{sec:limitations}

First, arXivDoc queries do not require multi-hop reasoning, and all queries are answerable. Each query is grounded in a single piece of evidence within one gold document. This keeps annotation reliable; however, whether our conclusions extend to settings requiring multi-hop reasoning or involving unanswerable queries remains untested.

Second, the corpus is only made of English arXiv paper and comparing with the ViDoRe (\S\ref{sec:vidore})
shows our conclusions do not transfer to that benchmark, so these findings
apply to text-rich scientific documents rather than document retrieval in general. Third,
the text\,+\,captions representation uses a single VLM captioner and prompt: changing either
 only minimally affects nDCG@10 (Appendix~\ref{app:caption_sensitivity}). Fourth, reported index sizes are
configuration-dependent — we did not sweep token pooling or dimensionality projection. We also did not report a budget-matched comparison at equal index size and latency — so the ratios in
Table~\ref{tab:main_results} are not architectural constants. Finally, we had only one annotator ran the full pipeline, with two others reviewing 50 sampled queries and agreeing with all
decisions, but we did not report inter-annotator agreement over the full set. Also, the figure subset
is small ($n=100$), with wide confidence intervals
(see Appendix~\ref{app:confidence_intervals}); and our false-negative check uses an LLM judge rather than a manual audit.

\section*{Ethics Statement}
We do not believe there are significant ethical issues associated with this research.


\bibliographystyle{colm2026_conference}

\appendix
\label{sec:appendix}
\section{Hyperparameters}
\label{app:hyperparameter}
Table~\ref{tab:retrieval_by_representation} reports retrieval accuracy (nDCG@10) under varying storage budgets, grouped by document representation. Storage is controlled through representation-specific hyperparameters, including text chunk size, visual token budgets (\texttt{max\_pixels}), and their combinations for interleaved representations.

\begin{table*}[h]
\centering
\small
\begin{tabular}{l l c c c c}
\toprule
\textbf{Representation} & \textbf{Model}  & \textbf{Index Size (GB)} &
\textbf{Text} & \textbf{Table} & \textbf{Figure} \\
\midrule

\multirow{6}{*}{Text}
& \multicolumn{4}{l}{\textit{Qwen}} \\
\cmidrule(l){2-5}
& chunk=512  & 2.56 & 0.87 & 0.60 & 0.76  \\
& chunk=1024  & 1.20 & 0.85 & 0.54 & 0.76 \\
& chunk=4096  & 0.34  & 0.83 & 0.47 & 0.64 \\
\cmidrule(l){2-5}
& \multicolumn{4}{l}{\textit{ColQwen}} \\
\cmidrule(l){2-5}
& chunk=1024  &  40.54 & 0.77 & 0.56 & 0.80 \\
& chunk=4096  & 40.54 & 0.77 & 0.57 & 0.80 \\

\midrule

\multirow{5}{*}{\LaTeX\ Figures}
& \multicolumn{4}{l}{\textit{ColQwen}} \\
\cmidrule(l){2-5}
& $\texttt{max\_pixels}=600K$   & 22.09 & 0.17 & 0.16 & 0.86 \\
&$\texttt{max\_pixels}=1M$   & 35.43 & 0.18 & 0.18 & 0.87 \\
& $\texttt{max\_pixels}=2M$   & 65.99 & 0.17 & 0.16 & 0.86  \\

\midrule

\multirow{5}{*}{Document-as-Image}
& \multicolumn{4}{l}{\textit{ColQwen}} \\
\cmidrule(l){2-5}
& $\texttt{max\_pixels}=1M$    & 47.23 & 0.71 & 0.51 & 0.83 \\
&$\texttt{max\_pixels}=2M$    & 94.79 & 0.73 & 0.52 & 0.84 \\

\bottomrule
\end{tabular}
\caption{Retrieval accuracy (NDCG@10) under varying storage budgets, grouped by document representation.}
\label{tab:retrieval_by_representation}
\end{table*}

\begin{figure*}[t]
\section{Query Generation}
\label{app:query_generation}

The query generation pipeline consists of four stages, three of which involve prompting \texttt{gpt-5.2}. In this section, we report the system prompts used at each stage.
\subsection{Synthetic Query Generation Prompt (Text)}
\centering
\begin{tcolorbox}[
    colback=gray!10!white,
    colframe=black!70!white,
    boxrule=0.4pt,
    width=\linewidth,
    arc=2mm,
    outer arc=2mm
]
\ttfamily
\footnotesize
\textbf{PROMPT}\vspace{0.2cm}

You are given extracted text from a scientific research paper. Your task is to generate a single, high-quality synthetic query that would meaningfully test a document retrieval system.\vspace{0.3cm}

\textbf{Instructions:}
\begin{enumerate}
    \item The query must require expert-level reasoning over implications, trends, limitations, or constraints discussed in the document, and must not be a direct restatement of any sentence from the input.
    \item The query must minimize lexical overlap with the input text by avoiding distinctive phrases or terminology, relying instead on abstraction and paraphrasing rather than keyword matching.
    \item The query must be answerable from the document but not trivially retrievable via keyword or phrase matching, and must not reference sections, figures, experiments, or document-specific wording.
    \item The query must ask exactly one focused question, without combining multiple sub-questions or enumerating parameters or conditions.
    \item The query must be realistic and concise, phrased as a single sentence that a knowledgeable researcher would plausibly ask, without verbose framing or artificial difficulty.
    \item If no query satisfying these criteria can be generated, return \texttt{null}.
\end{enumerate}

\vspace{0.3cm}
\textbf{Required Output Format:}

\begin{tcolorbox}[
    colback=white,
    colframe=black!40!white,
    boxrule=0.5pt,
    width=\linewidth,
    arc=2mm,
    outer arc=2mm
]
\ttfamily
\footnotesize
\begin{verbatim}
{
  "query": "<generated question or null>"
}
\end{verbatim}
\end{tcolorbox}

\vspace{0.3cm}
\textbf{Here is the document content:}

\texttt{\{paper\_text\}}
\end{tcolorbox}

\caption{Prompt used for generating synthetic, open-domain retrieval queries from scientific text.}
\label{fig:query_generation_prompt}
\end{figure*}

\begin{figure*}[t]
\subsection{Query Decontextualization Prompt}
\centering
\begin{tcolorbox}[
    colback=gray!10!white,
    colframe=black!70!white,
    boxrule=0.4pt,
    width=\linewidth,
    arc=2mm,
    outer arc=2mm
]
\ttfamily
\footnotesize
\textbf{PROMPT}\vspace{0.2cm}

You are a scientific question rewriter. You are given an original question that references a specific portion of a research paper. Your task is to rewrite it into a \textbf{context-independent, open-domain scientific query} that targets the same underlying concept, without relying on document-local or visual references.\vspace{0.3cm}

Do \textbf{not} refer to any figure, plot, panel, image, document, or use deictic expressions such as \emph{this}, \emph{that}, \emph{above}, or \emph{below}.\vspace{0.3cm}

\textbf{Requirements:}
\begin{enumerate}
    \item Preserve the core scientific intent, variables, and conditions present in the original question.
    \item Replace visual or deictic phrasing with concept-level wording (e.g., remove references such as ``based on the graph'' and ask directly about the relationship or effect).
    \item If symbols (e.g., $f_{\text{spec}}$) appear without definition, retain them exactly as written and do not invent meanings. A minimal parenthetical alias may be included only if it appears in the input.
    \item Remove all references to figures, plots, tables, panels, or document-local indices.
    \item Ensure the rewritten query can be answered by a knowledgeable reader without access to the original document or image.
    \item Retain units, ranges, and experimental or observational conditions if present.
    \item Avoid unresolved pronouns or placeholders (e.g., ``the parameter'', ``the system'') unless the domain makes them unambiguous.
    \item If the original question contains multiple sub-questions, keep only one and discard the rest.
    \item The final query must be a single, concise sentence with no superfluous framing or background.
\end{enumerate}

\vspace{0.3cm}
\textbf{Required Output Format:}

\begin{tcolorbox}[
    colback=white,
    colframe=black!40!white,
    boxrule=0.5pt,
    width=\linewidth,
    arc=2mm,
    outer arc=2mm
]
\ttfamily
\footnotesize
\begin{verbatim}
{
  "query": "<single rewritten question or null>",
  "reasoning": "<one-sentence rationale>"
}
\end{verbatim}
\end{tcolorbox}

\vspace{0.3cm}
\textbf{If a valid context-independent query cannot be produced, set \texttt{"query"} to \texttt{null} and briefly explain why in \texttt{"reasoning"}.}
\end{tcolorbox}

\caption{Prompt used for decontextualizing document-dependent scientific questions into open-domain queries.}
\label{fig:query_decontextualization_prompt}
\end{figure*}

\begin{figure*}[t]
\subsection{Query Verification Prompt}
\centering
\begin{tcolorbox}[
    colback=gray!10!white,
    colframe=black!70!white,
    boxrule=0.4pt,
    width=\linewidth,
    arc=2mm,
    outer arc=2mm
]
\ttfamily
\footnotesize
\textbf{PROMPT}\vspace{0.2cm}

You are a validator that checks whether a \emph{decontextualized question} is well-formed for open-domain retrieval.  
Judge \emph{only} from the provided JSON fields. Do \textbf{not} assume access to the original figure, table, or paper.\vspace{0.3cm}

\textbf{What ``valid decontextualized question'' means:}

A question is \textbf{valid} \emph{if and only if} all of the following criteria are satisfied:
\begin{enumerate}
    \item \textbf{Context-independent}: The question contains no references to local context such as ``this figure,'' ``the table above,'' ``these results,'' or any indexical phrasing that requires the original document or image.
    \item \textbf{Answerable in principle}: A knowledgeable person or external source could answer the question without access to the original paper or figure. The domain and variables must be sufficiently specified. Crucially, the question must not rely on parameters, symbols, or notations that are defined arbitrarily or only within the source paper (e.g., a tuning parameter with no standard meaning in the field).
    \item \textbf{Intent preserved}: The question targets the same underlying information need as the original question, but generalized beyond the local figure or document context.
    \item \textbf{Clarity and unambiguous entities}: Any entities, variables, or notations must be interpretable by an expert in the relevant field without requiring the specific paper. Unresolved pronouns or placeholders (e.g., ``the parameter,'' ``the system'') are not allowed unless they are standard and unambiguous within the domain.
\end{enumerate}

\vspace{0.3cm}
\textbf{Guiding Principle for Ambiguity:}

Requiring background domain knowledge is acceptable and expected for real search queries. However, ambiguity arising from terms that are defined only within the source document or that depend on the original figure context is not acceptable.\vspace{0.3cm}

\textbf{Common failure modes (label them if present):}
\begin{itemize}
    \item \texttt{underspecified\_parameter} (especially if defined arbitrarily in the source paper)
    \item \texttt{still\_context\_bound}
    \item \texttt{domain\_missing\_or\_vague}
    \item \texttt{ambiguity\_pronouns\_placeholders}
    \item \texttt{unanswerable\_generic}
\end{itemize}

\vspace{0.3cm}
\textbf{Required Output Format (JSON only):}

\begin{tcolorbox}[
    colback=white,
    colframe=black!40!white,
    boxrule=0.5pt,
    width=\linewidth,
    arc=2mm,
    outer arc=2mm
]
\ttfamily
\footnotesize
\begin{verbatim}
{
  "is_valid": boolean,
  "score": integer,             
  "decision_rationale": string,
  "confidence": integer
}
\end{verbatim}
\end{tcolorbox}

\end{tcolorbox}

\caption{Prompt used to verify whether a generated question is a valid, context-independent query suitable for open-domain retrieval.}
\label{fig:query_verification_prompt}
\end{figure*}

\begin{figure*}[t]
\section{Example Queries}
\label{app:query_examples}
\centering
\begin{tcolorbox}[
    colback=yellow!8!white,
    colframe=black!70!white,
    boxrule=0.4pt,
    width=\linewidth,
    arc=2mm,
    outer arc=2mm
]
\ttfamily
\footnotesize

\textbf{arXiv ID:} \texttt{1007.4239}\\
How can an optical system rapidly and cheaply switch between four orbital angular momentum (helical-phase) mode indices while keeping polarization independent?

\vspace{0.3cm}
\textbf{arXiv ID:} \texttt{1011.0302}\\
In the dense phase of the O(n) loop model, how does changing the topology of the graph representation determine if a local perturbation is RG-relevant or RG-irrelevant according to Coulomb gas scaling dimensions?

\vspace{0.3cm}
\textbf{arXiv ID:} \texttt{1806.03783}\\
How does increasing the normalized accretion rate lead to a turnover in the fraction of IR luminosity reprocessed by circumnuclear dust?

\vspace{0.3cm}
\textbf{arXiv ID:} \texttt{1205.2806}\\
Which ion–neutral atom pair best enables quantum-threshold scattering while being least sensitive to stray static electric fields, i.e., with a high threshold energy but low static-field strength needed for excess micromotion to reach that energy?

\vspace{0.3cm}
\textbf{arXiv ID:} \texttt{1304.2695}\\
How do the implicit midpoint and trapezoidal (Crank–Nicolson) rules differ in using a single midpoint evaluation versus averaging endpoint evaluations?

\vspace{0.3cm}
\textbf{arXiv ID:} \texttt{1411.3004}\\
Which subtype of pulsating star shows the strongest positive correlation between variability amplitude and pulsation period across stars spanning a wide range of distances?

\vspace{0.3cm}
\textbf{arXiv ID:} \texttt{2006.00262}\\
What is the sequence of processing steps that transforms source and target monolingual corpora into mapped cross-lingual word embeddings (CLWEs)?

\end{tcolorbox}

\caption{Representative examples of decontextualized, evidence-grounded queries in ArXivDoc. Each query targets a specific piece of scientific evidence.}
\label{fig:query_examples}
\end{figure*}

\begin{figure*}[t]
\subsection{Human Annotation Protocol}
\centering
\begin{tcolorbox}[
    colback=gray!10!white,
    colframe=black!70!white,
    boxrule=0.4pt,
    width=\linewidth,
    arc=2mm,
    outer arc=2mm
]
\ttfamily
\footnotesize
\textbf{ANNOTATION GUIDELINES}\vspace{0.2cm}

You are evaluating scientific retrieval queries generated from research documents. Queries may be questions or short search-style phrases. Each query must be assessed and, if necessary, revised based on three criteria: \textbf{naturalness}, \textbf{ambiguity}, and \textbf{document answerability}.\vspace{0.3cm}

\textbf{Criteria:}
\begin{enumerate}
    \item \textbf{Naturalness:} The query must be clear, understandable, and plausible as a realistic search query. A query is considered unnatural if it (i) combines multiple questions (e.g., ``What is X, and why Y?''), (ii) contains excessive domain-specific terminology (approximately 8 or more specialized terms), (iii) is overly verbose (e.g., more than three clauses), or (iv) is difficult to interpret. Such queries should be rewritten to improve clarity while preserving the original scientific intent.

    \item \textbf{Ambiguity:} The query must be specific enough to identify a single target document. A query is considered ambiguous if it is overly broad, underspecified, admits multiple interpretations, or could be answered by many documents. Ambiguous queries should be rewritten to resolve underspecification and better target the intended document, and if that's not a possibility, they should be removed.

    \item \textbf{Document Answerability:} The query must be directly supported by evidence in the document. The annotator must verify that a specific passage, parsed table, or rendered figure from the \LaTeX{} source contains the information needed to answer the query. Queries whose premise is unsupported or contradicted by the document, or for which no identifiable evidence exists, must be removed.
\end{enumerate}

\vspace{0.3cm}
\textbf{Workflow:}
\begin{enumerate}
    \item Perform initial evidence screening using retrieval tools (e.g., NotebookLM or Gemini Flash 2.5) to surface candidate passages, tables, or figures and identify potential issues with naturalness, ambiguity, or answerability. 
    \item Assign a coarse quality score (1--10) to guide assessment of ambiguity and retrieval specificity; this score is used for calibration and not thresholded directly. 
    \item Manually inspect the retrieved evidence against the original \LaTeX{} source, including text, parsed tables, and rendered figures, to confirm correctness.
    \item Rewrite queries to improve clarity and specificity while preserving intent, or discard queries that cannot be made valid.
\end{enumerate}

\end{tcolorbox}
\caption{Human annotation protocol for evaluating and refining scientific retrieval queries.}
\label{fig:human_annotation}
\end{figure*}
\clearpage

\subsection{Caption Generation and Sensitivity}
\label{app:caption_sensitivity}

Figure captions for the text\,+\,VLM captions representation were generated with
\texttt{gpt-5-mini} at default decoding settings (temperature $=1$), using the short
prompt in Figure~\ref{fig:prompt_caption_short}. The prompt was written once and applied
unchanged to all $8{,}210$ documents; we did not perform any prompt engineering.
Captioning the full corpus cost approximately \$200.

The representation therefore depends on two choices: which model generates the captions,
and what prompt it is given. To test whether our results depend on either, we regenerate
captions for the full corpus under two variations, holding the retrieval model
(Qwen3-Embedding) and all other settings fixed. The first substitutes an open-weight
captioner, Llama 4 Maverick, using the same short prompt. The second keeps
\texttt{gpt-5-mini} but replaces the prompt with a substantially richer,
schema-constrained variant (Figure~\ref{fig:prompt_caption_rich}) that asks for verbatim
axis labels, legend entries, table headers, and named entities.

\begin{table}[h]
\centering
\small
\begin{tabular}{lllc}
\toprule
\textbf{Captioner} & \textbf{Prompt} & \textbf{Variation} & Avg. nDCG@10 \\
\midrule
\texttt{gpt-5-mini} & short & default (main results)    & 0.75 \\
\texttt{Llama 4 Maverick}    & short & open-weight captioner     & 0.75 \\
\texttt{gpt-5-mini} & rich  & schema-constrained prompt & 0.76 \\
\bottomrule
\end{tabular}
\caption{Sensitivity of the text\,+\,VLM captions representation to captioner and prompt,
averaged over text, table, and figure queries (Qwen3-Embedding). Neither swapping in an
open-weight captioner nor using a substantially richer prompt changes performance by more
than $0.01$.}
\label{tab:caption_sensitivity}
\end{table}

Neither variation changes the result by more than $0.01$, indicating that the
representation does not depend on a particular captioner or on prompt tuning. We note
that all three settings produce a static text description per figure, so this experiment
varies caption \emph{quality} rather than the captioning objective itself; we discuss the
resulting limitation in \S\ref{sec:limitations}.

\begin{figure*}[h]
\centering
\begin{tcolorbox}[
    colback=gray!10!white,
    colframe=black!70!white,
    boxrule=0.4pt,
    width=\linewidth,
    arc=2mm,
    outer arc=2mm
]
\ttfamily
\footnotesize
\textbf{SHORT PROMPT (DEFAULT)}\vspace{0.2cm}

Analyze the image and briefly describe what it shows, including the names of visible
elements (e.g., variables, axes, legends, tables, curves, diagram components) using the
following JSON structure:\vspace{0.2cm}

\{\\
\hspace*{1em}"explanation": "A concise description of the image, naming key elements that
appear and describing their relationships only at a high level"\\
\}
\end{tcolorbox}
\caption{Short prompt used to generate figure captions for the text\,+\,VLM captions
representation in our main results.}
\label{fig:prompt_caption_short}
\end{figure*}

\begin{figure*}[h]
\centering
\begin{tcolorbox}[
    colback=gray!10!white,
    colframe=black!70!white,
    boxrule=0.4pt,
    width=\linewidth,
    arc=2mm,
    outer arc=2mm
]
\ttfamily
\footnotesize
\textbf{RICH PROMPT (SENSITIVITY CHECK)}\vspace{0.2cm}

Produce a thorough description that lets a downstream text retriever surface this image
for queries about its content. Be specific and verbatim where possible: copy axis labels,
legend entries, table headers, row/column names, units, equation variables,
dataset/method names, and any title or caption-like text exactly as written.\vspace{0.2cm}

Required JSON schema:\vspace{0.2cm}

\{\\
\hspace*{1em}"figure\_type": "One of: line\_plot | bar\_chart | scatter | heatmap |
histogram | box\_plot | violin\_plot | pie\_chart | flow\_diagram |
architecture\_diagram | schematic | algorithm\_block | table | equation | screenshot |
photograph | map | mixed | other",\\
\hspace*{1em}"caption": "A 1-3 sentence high-level summary describing what the image shows
and the main finding/relationship it conveys.",\\
\hspace*{1em}"axes\_or\_columns": "For plots, list axes with their labels, units, and
ranges.",\\
\hspace*{1em}"key\_entities": ["Verbatim list of named variables, methods, models,
datasets, conditions."],\\
\hspace*{1em}"numerical\_or\_textual\_content": "Notable numbers, percentages, p-values,
statistics, equation expressions",\\
\hspace*{1em}"visual\_relationships": "1-3 sentences describing trends",\\
\hspace*{1em}"context\_signals": "Any visible hints about the scientific subdomain or
task."\\
\}
\end{tcolorbox}
\caption{Richer, schema-constrained prompt used in the captioner-sensitivity experiment
(Table~\ref{tab:caption_sensitivity}). It is not used in the main results.}
\label{fig:prompt_caption_rich}
\end{figure*}
\clearpage
\subsection{Confidence Intervals}
\label{app:confidence_intervals}

ArXivDoc contains $547$ queries, and the figure-based subset in particular is small
($n=100$). To quantify how much of the difference between representations is resolvable
at this scale, we compute nDCG@10 with bootstrapped $95\%$ confidence intervals, using
$10{,}000$ resamples drawn with replacement over queries within each query type.

\begin{table}[H]
\centering
\small
\begin{tabular}{lcc}
\toprule
\textbf{Query type} & \textbf{Qwen3-Embedding (Text-Only)} & \textbf{ColQwen (Doc-as-Image)} \\
\midrule
Text (\textit{n}=229)    & 0.87 [0.85, 0.89] & 0.74 [0.70, 0.78] \\
Table (\textit{n}=218)   & 0.60 [0.55, 0.65] & 0.52 [0.47, 0.57] \\
Figure (\textit{n}=100)  & 0.76 [0.68, 0.83] & 0.84 [0.78, 0.90] \\
\midrule
Overall (\textit{n}=547) & 0.75 [0.72, 0.77] & 0.67 [0.64, 0.70] \\
\bottomrule
\end{tabular}
\caption{nDCG@10 with bootstrapped $95\%$ confidence intervals ($10{,}000$ resamples).
Intervals are non-overlapping for overall, text, and table queries, and overlap on the
figure subset.}
\label{tab:ci}
\end{table}

The intervals for overall, text, and table queries do not overlap, so the advantage of
text-based representations on those subsets is resolvable at this sample size. On figure
queries the intervals overlap, and the subset is small enough that intervals span
$0.15$ in width; differences on figure queries should therefore be read as suggestive
rather than established. This is the main respect in which the benchmark's scale limits
what can be concluded, and expanding the figure-query subset is a natural direction for
future versions of ArXivDoc.

\end{document}